\documentclass[letterpaper,11pt,twoside]{article}
\usepackage{amsmath,amssymb,amscd}
\usepackage{color}
\usepackage[dvips]{graphicx}

\newcommand{\im}{\mathrm{i}}  
\newcommand{\defeq}{:=}
\newcommand{\C}{\mathbb{C}}
\newcommand{\tens}{\otimes}   
\newcommand{\xd}{\mathrm{d}}
\newcommand{\xD}{\mathcal{D}} 
\DeclareMathOperator{\id}{id}
\newcommand{\cH}{\mathcal{H}}
\newcommand{\cHo}{\mathcal{H}_{\textrm{O}}}
\newcommand{\cHc}{\mathcal{H}_{\textrm{C}}}
\newcommand{\cHob}{\mathcal{H}_{\bar{\textrm{O}}}}
\newcommand{\cHcb}{\mathcal{H}_{\bar{\textrm{C}}}}

\newcommand{\tauo}{\tau_\textrm{OO}}
\newcommand{\tauc}{\tau_\textrm{OC}}
\newcommand{\iotao}{\iota_\textrm{O}}
\newcommand{\iotac}{\iota_\textrm{C}}
\DeclareMathOperator{\tr}{tr}
\newcommand{\falg}{\mathcal{C}}  
\newcommand{\falgc}{\mathcal{C}_\textrm{class}}  

\begin{document}

\begin{titlepage}
\title{\textbf{Two-dimensional quantum Yang-Mills theory with corners}}
\author{Robert Oeckl\footnote{email: robert@matmor.unam.mx}\\ \\
Instituto de Matem\'aticas, UNAM, Campus Morelia,\\
C.P. 58190, Morelia, Michoac\'an, Mexico}
\date{UNAM-IM-MOR-2006-2\\ 30 August 2006\\ 17 July 2007 (v2)}

\maketitle

\vspace{\stretch{1}}

\begin{abstract}
The solution of quantum Yang-Mills theory on arbitrary compact
two-manifolds is well known. We bring this solution into a TQFT-like
form and extend it to include corners. Our formulation is based on an
axiomatic system that we hope is flexible enough to capture actual
quantum field theories also in higher dimensions. We motivate this
axiomatic system from a formal Schr\"odinger-Feynman quantization
procedure. We also discuss the physical meaning of unitarity, the
concept of vacuum, (partial) Wilson loops and non-orientable surfaces.

\end{abstract}

\vspace{\stretch{1}}
\end{titlepage}

\section{Introduction}

The subject of two-dimensional quantum Yang-Mills theory is an old
one. Solvability of the theory was already shown by Migdal
\cite{Mig:reclgt}, using the lattice approach. The advent of
topological quantum field theory (TQFT) and related ideas generated
interest in this
two-dimensional theory from a new perspective. The theory was
formulated and solved on arbitrary (compact) two-manifolds with
boundaries \cite{Wit:qgauge2d,Fin:qymriem,Rus:gauge2d,BlTh:qym}.
In the present paper we wish to carry this one step further, namely by
allowing generalized manifolds which may have corners. Roughly, this
means that the boundaries are not necessarily closed, but may have
boundaries themselves.
While there is already considerable work on TQFT with corners, usually
following Walker \cite{Wal:tqftnotes}, this
generally does not extend to the situation where the vector spaces
associated with boundaries are infinite dimensional.
We show that two-dimensional quantum Yang-Mills theory provides
a realization for a TQFT-type system of axioms that admits both
infinite dimensionality of vector spaces and manifolds with corners.

Another motivation for the present work comes from the general
boundary formulation of quantum mechanics
\cite{Oe:catandclock,Oe:boundary,CDORT:vacuum,Oe:GBQFT}.
The
inclusion of corners in this framework, which is based on a specific
TQFT-type system of axioms is an outstanding
problem, see the discussion in \cite{Oe:GBQFT}. Two-dimensional
quantum Yang-Mills theory being an actual
quantum mechanical system should thus serve as an example of a
``physically correct'' implementation of corners. In
particular, this implementation must be compatible with the probability
interpretation as outlined in \cite{Oe:GBQFT}.
Indeed, the version of two-dimensional quantum Yang-Mills theory
constructed in this paper is unitary in the extended sense of
\cite{Oe:GBQFT} and hence compatible with this probability
interpretation. In particular, we show how corners play a role in
the deformation of regions and allow to formulate the associated
probability conservation condition.

The axioms presented in this paper are a direct generalization of those
presented in \cite{Oe:GBQFT}. We comment both on the
mathematical as well as the physical motivation for the specific type
of generalization we perform. As in the case without corners, the
physical justification for the axioms comes from a simple
quantization prescription, combining the Schr\"odinger representation
with the Feynman path integral.

Additional topics we cover are the axiomatization and realization of
the concept of vacuum, the inclusion of Wilson loops and the extension
to non-orientable manifolds.

Section~\ref{sec:axioms} is concerned with the axiomatic system, its
mathematical and physical motivations, and a comparison to the
axiomatic system of \cite{Oe:GBQFT}. Section~\ref{sec:2dax} elaborates
on the two-dimensional case, identifying elementary data in this
context. Section~\ref{sec:2dqym} then develops two-dimensional quantum
Yang-Mills theory as a realization of the
axioms. Section~\ref{sec:extensions} extends this by including the
concept of vacuum, Wilson loops and non-orientable surfaces. A closing
section presents a brief outlook.

\section{The axiomatic system}
\label{sec:axioms}

\subsection{Mathematical motivation}

We provide motivations for a specific set of axioms (to be introduced
subsequently) that might broadly be identified as describing a type of
topological quantum field theory.
Note that the attribute ``topological''
does not necessarily mean that we consider topological manifolds only,
although we shall initially do so.

Recall the basic setup of a topological quantum field
theory (TQFT) \cite{Ati:tqft}.
We associate finite dimensional vector spaces
$\cH_\Sigma$ with
$(n-1)$-dimensional manifolds $\Sigma$ and maps between these vector
spaces with $n$-dimensional
cobordisms. An $n$-dimensional cobordism is an
$n$-dimensional manifold $M$ with boundary $\Sigma$ so that the
boundary is the disjoint union of two $(n-1)$-manifolds $\Sigma =
\Sigma_1\cup \Sigma_2$. Thus we associate with $M$ a linear map
$\cH_{\Sigma_1}\to\cH_{\Sigma_2}$, declaring $\Sigma_1$ to be the
``in''-component and $\Sigma_2$ to be the ``out''-component of the
boundary. A key requirement is then that the
map associated with a cobordism that arises as the gluing of two
cobordisms is the composition of the maps associated with the glued
cobordisms. This allows a functorial formulation of TQFT, i.e., as a
functor from the category of $(n-1)$-manifolds and $n$-cobordisms to
the category of finite dimensional vector spaces and linear maps.

This manner of axiomatizing TQFT faces limitations once we wish to
consider infinite dimensional vector spaces (as becomes necessary if
we want to describe real quantum field theories). For example, let $n=2$
and consider a cylinder as a cobordism between two circles. Associated
to this is a map from the vector space for a circle to itself. This
map is in fact the identity.\footnote{Note that this does not
follow from what we have said so far. In general this map would be a
projector. However, without loss of information we might restrict the
state space to its domain, making it the identity. This is generally
done in TQFT.} Hence, gluing the two
circles together to form a torus yields the trace of the identity
map. This is the dimension of the vector space and hence infinite if
it is infinite dimensional. We could avoid this kind of problem by
restricting the class of admissible closed $n$-manifolds.

However,
a related problem cannot be eliminated in this way. Namely, in
general there are many ways to arrange the connected components of the
boundary of an
$n$-manifold into an ``in''- and an ``out''-boundary. Not all of these
would generally lead to well defined maps. In particular, choosing the
whole boundary to be ``out'' will generally not lead to a well defined
map (due to the same type of infinities
as above). We avoid this problem by always taking the whole
boundary to be ``in''. Of course this means that we have to
reformulate the correspondence between gluing and composition, loosing
its simple functorial formulation.

We also require both the $n$-manifolds and
$(n-1)$-manifolds under consideration to be
oriented. Furthermore, reversal of orientation of an $(n-1)$-manifold
corresponds to dualization of the associated vector space. Hence, we
want the bi-dual space to be isomorphic to the original one. To
achieve this in the infinite dimensional situation we add structure
and consider Hilbert spaces. The use of Hilbert spaces has another
essential reason, namely the applicability of the physical
interpretation in terms of quantum mechanics and probabilities.

The points discussed so far are implemented in the axiomatic
system presented in
\cite{Oe:GBQFT} together with an extended quantum mechanical probability
interpretation. Furthermore, it was shown in
\cite{Oe:timelike,Oe:KGtl} that the
Klein-Gordon quantum field theory
satisfies these axioms and the associated physical interpretation at
least for certain special classes of manifolds.
A crucial element that is missing so far, but is desirable
from a physics point of view (see the discussion in \cite{Oe:GBQFT})
are \emph{corners}.

Let us mention first that the manifolds in question may carry
structure in addition to being topological manifolds. For the moment
we shall only be interested in the case of no additional structure and
that of differentiable structure.
Before we proceed we change our terminology slightly: In the following
we refer to the oriented
cobordisms or $n$-manifolds as \emph{regions} and to the oriented
$(n-1)$-manifolds as \emph{hypersurfaces}.

\begin{figure}[h]
\begin{center}
\input{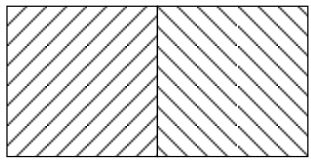}
\end{center}
\caption{Gluing two rectangles to form a new rectangle.}
\label{fig:rectglue}
\end{figure}

Now imagine that we want to glue two regions with the shape of solid
rectangles along one side to form a new region with the shape of a
solid rectangle, see Figure~\ref{fig:rectglue}. This situation is not
covered by the standard TQFT
axioms. Consider first the differentiable situation: The regions in
question are not even differentiable manifolds. To admit them we have
to generalize the definition of a region. Rather than being manifolds
with boundary they will be manifolds with boundaries and
\emph{corners}. In the present example it is quite clear what this
means, but we deliberately avoid a precise general definition
here. On the other hand, the boundary of a
rectangle is still a valid differentiable 1-manifold and seen as a
part of it the
corners become ``invisible''. However, they do play a role in the
gluing as only a part of the boundary is glued. Moreover, this part is
not a connected component as required by standard axioms. Rather it is
itself bounded by two corners. If we consider topological manifolds,
the rectangles are homeomorphic to discs and as such perfectly well
defined topological 2-manifolds with boundary. However, the corners
still make their appearance as boundaries of parts of boundaries along
which regions are glued. Indeed, without them it would be impossible
to glue two discs to a disc.

We will refer to corners that are not explicitly visible in the
differentiable or topological structure as \emph{virtual
corners}. Hence, in a
topological context, to which we restrict in the following, all
corners in the boundaries of regions are virtual. We try to implement
them in a minimalistic
fashion, complicating the axioms as little as possible. To this end we
introduce a concept of \emph{decomposition} of a hypersurface into
hypersurfaces as follows. (One might conversely think of this as
providing a notion of gluing of hypersurfaces.) Hypersurfaces are thus
oriented topological manifolds of dimension $n-1$ with boundary, and
regions are oriented topological manifolds of dimension $n$ with
boundary.
A decomposition of a
hypersurface $\Sigma$ is the
presentation of $\Sigma$ as a finite
union of hypersurfaces $\Sigma_1,\dots,\Sigma_n$ with the following
properties. Each $\Sigma_i$ is closed in $\Sigma$ and the intersection
of any $\Sigma_i$ with any $\Sigma_j$ is the intersection
of their boundaries. Note that this definition of decomposition
includes as a special case a decomposition into disjoint components.
In addition to the manifest corners of a hypersurface, namely its
boundary, a decomposition provides additional virtual corners. These
are the boundaries of the pieces of the decomposition which are not
already contained in the boundary of the original hypersurface.

Before presenting the generalization of the system of axioms of
\cite{Oe:GBQFT}, we mention another modification which was already
discussed in \cite{Oe:GBQFT}. Namely, to simplify the axioms it is
convenient to introduce \emph{empty regions}. Essentially, these are
oriented topological manifolds of dimension $n-1$ with boundary, as
are the
hypersurfaces. However, they should be thought of as regions
completely contracted to their boundary. Hence, the boundary of an
empty region is defined to be the union of two copies of it as a
hypersurface, but with opposite orientations. Furthermore, these
copies are glued along their boundaries, providing a hypersurface with
decomposition.

\subsection{Core axioms}
\label{sec:axiomlist}

We are now ready to list the axioms. If a hypersurface is denoted by
$\Sigma$, its oppositely oriented version is denoted by $\bar{\Sigma}$.

\begin{itemize}
\item[(T1)] Associated to each hypersurface $\Sigma$ is a complex
  separable Hilbert space $\cH_\Sigma$, called the \emph{state space} of
  $\Sigma$. We denote its inner product by
  $\langle\cdot,\cdot\rangle_\Sigma$.
\item[(T1b)] Associated to each hypersurface $\Sigma$ is an antilinear
  isomorphism $\iota_\Sigma:\cH_\Sigma\to\cH_{\bar{\Sigma}}$. This map
  is an   involution in the sense that
  $\iota_{\bar{\Sigma}}\circ\iota_\Sigma$ is the identity on
  $\cH_\Sigma$.
\item[(T2)] Suppose the hypersurface $\Sigma$ decomposes into a union
  of hypersurfaces $\Sigma_1\cup\cdots\cup\Sigma_n$. Then, there is a
  bounded surjective map of state spaces
  $\tau:\cH_{\Sigma_1}\tens\cdots\tens\cH_{\Sigma_n}\to\cH_\Sigma$.
  Furthermore, the restriction of $\tau$ to the orthogonal complement
  of its kernel preserves the inner product, i.e., is an isomorphism
  of Hilbert spaces. If
  the decomposition is disjoint $\tau$ is also injective. The
  composition of the maps $\tau$ associated with two consecutive
  decompositions is identical to the map $\tau$ arising from the
  resulting decomposition.
\item[(T2b)] The involution $\iota$ is compatible with the above
  decomposition. That is,
  $\tau\circ(\iota_{\Sigma_1}\tens\cdots\tens\iota_{\Sigma_n}) 
  =\iota_\Sigma\circ\tau$.
\item[(T4)] Associated with each region $M$ is a linear map
  from the state space of its boundary $\Sigma$ (with induced
  orientation) to the complex
  numbers, $\rho_M:\cH_\Sigma\to\C$. This is called the
  \emph{amplitude} map.
\item[(T3x)] Suppose $M$ is an empty region. Then its boundary
  can be decomposed into two components that are identical
  up to orientation, $\bar{\Sigma}\cup\Sigma$. The
  extended amplitude map $(\cdot,\cdot)_\Sigma\defeq\rho_M\circ\tau$
  defines a bilinear pairing $\cH_{\bar{\Sigma}}\tens\cH_{\Sigma}\to\C$.
  We require this pairing to be compatible with the involution and
  Hilbert space structure in the sense that
  $\langle\cdot,\cdot\rangle_\Sigma=(\iota_\Sigma(\cdot),\cdot)_\Sigma$.
\item[(T4b)] Suppose $M$ is a region with boundary $\Sigma$,
  decomposable into the union of two components,
  $\Sigma=\Sigma_1\cup\Sigma_2$. Suppose the extended amplitude map
  $\rho_M\circ\tau:\cH_{\Sigma_1}\tens\cH_ {\Sigma_2}\to\C$ gives rise
  to an isomorphism of vector spaces
  $\tilde{\rho}_M:\cH_{\Sigma_1}\to\cH_{\bar{\Sigma}_2}$. Then we require
  $\tilde{\rho}_M$ to preserve the inner product, i.e., be
  \emph{unitary}.
\item[(T5)] Let $M_1$ and $M_2$ be two regions such that the union
  $M=M_1\cup M_2$ is again a region and the intersection is a
  hypersurface $\Sigma$. The boundary of $M_1$ may be decomposed into
  $\Sigma_1\cup\Sigma$ and the boundary of $M_2$ into
  $\Sigma_2\cup\bar{\Sigma}$. Let $\{\xi_i\}_{i\in I}$ be an ON-basis
  of $\cH_{\Sigma}$. If
\[
\sum_{i\in I}\rho_{M_1}\circ\tau_1(\cdot\tens\xi_i)\,
\rho_{M_2}\circ\tau_2(\cdot\tens\xi_i^*)
\]
exists then we require it to be equal to $\rho_M\circ\tau(\cdot\tens\cdot)$.
\end{itemize}

The strange seeming numbering of the axioms is provided merely for easier
comparison to \cite{Oe:GBQFT}. Let us briefly perform such a
comparison. As already mentioned, we admit empty regions here which
allows us to remove axioms (T3) and (T3b) of \cite{Oe:GBQFT} and
replace them with a single axiom (T3x). This appears now after (T4),
however, as it requires the amplitude map to be defined. A further
change is the formulation of the gluing (T5) via the
insertion of an ON-basis instead of a functorial formulation
(conditional on existence). This avoids the artificial introduction of
an ``in''/``out'' splitting of boundaries.
We also require consecutive
decompositions of hypersurfaces to yield the same map $\tau$ as the
resulting decomposition. While this was implicitly understood it is
now written explicitly in (T2).

We now turn to the modification of the axioms mandated by the
implementation of corners. This modification is effected by replacing
the concept of decomposition of hypersurfaces from the more special
one admitting only decompositions into connected components to the
more general one defined above. Hence, the
modification of the axioms is mostly implicit. This is the case for
axioms (T2b), (T4b) and (T5). The single axiom where the
implementation of corners is more explicit is axiom (T2). Here, we
encounter the novel possibility that the map $\tau$ associated with a
decomposition is not generally an isomorphism. Indeed, it is only an
isomorphism in the case that the decomposition is into disjoint
components.

If we restrict the concept of decomposition to the disjoint one, we
recover the old axioms (although with the changes unrelated to corners
mentioned above). In
that case (T2) simplifies a little bit, but no other explicit change
appears. (One might remove the explicit mention of $\tau$ in (T4b) and
(T5), but this is more a matter of aesthetics.)
In this sense our implementation of corners may be said to be
minimalistic. Of course, another implicit change is in the
definition of hypersurfaces. In the case without corners we
would not admit hypersurfaces to have boundaries.

Besides mathematical minimalism our proposed implementation of corners
is motivated from the physical requirement to match actual quantum
field theories. In this direction our proposal arises from a
Schr\"odinger-Feynman quantization prescription. This is the subject
of the next section.

\subsection{Schr\"odinger-Feynman quantization and corners}
\label{sec:quantize}

Recall that topological quantum field theory, even though it may be
considered a purely mathematical subject, arose out of methods
of quantum field theory and conformal field theory \cite{Seg:cftdef}.
Roughly speaking, the vector spaces associated with boundaries were
thought of as analogs of state spaces of a quantum mechanical
system, while the
maps between these vector spaces were thought of as analogs
of time evolution operators.

The
axioms presented in \cite{Oe:GBQFT} and refined in
the previous section may be seen as a late attempt to bring these
ideas back into physics, by which we mean here to ordinary
quantum field theory. One part of this endeavor is to develop the
formal framework,
i.e., the set of axioms together with their physical
interpretation. Another part of this is to provide
concrete theories fitting the framework. Usually, quantum theories
are obtained through a process of quantization. Unfortunately, at
present there exists no fully satisfactory quantization prescription
for the framework considered here. However, a Schr\"odinger-Feynman
quantization (which motivated TQFT originally) works on a formal level
and can be made to work at least in some
situations of physical interest \cite{Oe:boundary,Oe:timelike,Oe:KGtl}.
We recall its most essential elements from the perspective of the
application to our axiomatic system. We give special emphasis to the
implementation of corners in this context.
For more details (without
corners) we refer the reader to \cite{Oe:GBQFT} or \cite{Oe:KGtl}.

Consider a classical field theory. Hence,
a space $K_\Sigma$ of (field) configurations is associated with each
hypersurface $\Sigma$ together with a measure on this configuration
space. We define the state space $\cH_\Sigma$ as the space of complex
square integrable functions, called \emph{wave functions}, on
$K_\Sigma$ with the inner product
\begin{equation}
  \langle\psi,\psi'\rangle_\Sigma\defeq\int_{K_\Sigma}\xD\varphi\,
  \overline{\psi(\varphi)}\psi'(\varphi) .
\end{equation}
The involution of axiom (T1b) is simply the complex conjugation of
wave functions,
\begin{equation}
 (\iota_\Sigma(\psi))(\varphi)\defeq\overline{\psi(\varphi)}\quad\forall
 \psi\in\cH_\Sigma,\varphi\in K_\Sigma .
\label{eq:conj}
\end{equation}

Suppose we have a decomposition of a hypersurface $\Sigma$ into two
components, $\Sigma_1$ and $\Sigma_2$. This yields an injective map
$K_\Sigma\to K_{\Sigma_1}\times K_{\Sigma_2}$ between the associated
state spaces by simply
forgetting parts of the configuration data. (This supposes the
configuration data to be local in a suitable sense.) Hence, we obtain
an induced surjective linear map
$\tau:\cH_{\Sigma_1}\tens\cH_{\Sigma_2}\to\cH_\Sigma$ between the
associated spaces of wave functions.
Explicitly,
\begin{equation}
(\tau(\psi\tens\eta))(\varphi)=
\psi(\varphi|_{\Sigma_1})\eta(\varphi|_{\Sigma_2})
\quad\forall\psi\in\cH_{\Sigma_1},\eta\in\cH_{\Sigma_2},
 \varphi\in K_{\Sigma} .
\label{eq:tauphi}
\end{equation}
If the decomposition is disjoint,
no configuration data is ``forgotten'' and the maps are
bijective. Indeed, the latter is then an isomorphism of Hilbert spaces
since the inner product on the tensor product is induced from the
inner products on the components.
The latter case provided the motivation for the old version of axiom
(T2), which is standard in TQFT, while the new version is motivated by
the general case with
corners implemented through generalized decompositions.

The amplitude (T4) for a region $M$ is given by an integral of the
wave function over boundary configurations weighted by a kernel $Z_M$,
called the \emph{field propagator}.
\begin{gather}
 \rho_M(\psi)\defeq\int_{K_\Sigma}\xD\varphi\, \psi(\varphi) Z_M(\varphi)
 \quad\forall \psi\in\cH_\Sigma,
\label{eq:ampl}
\end{gather}
The field propagator in turn is defined through an action $S_M$,
which is a function on a space of configurations $K_M$ on $M$,
\begin{gather}
 Z_M(\varphi)\defeq\int_{K_M, \phi|_\Sigma=\varphi}\xD\phi\, e^{\im
 S_M(\phi)}\quad\forall \varphi\in K_\Sigma .
\label{eq:prop}
\end{gather}

Supposing the above definitions can be made rigorous, the axioms are
then automatically satisfied (except for unitarity).
In particular, axiom (T5) follows from formal gluing properties of the path
integral. In the case without corners this is detailed in
\cite{Oe:GBQFT} and \cite{Oe:KGtl}. In the case with corners,
for axiom (T2) this was
explained above. The only other axiom were the corners make an
essential difference is (T5). Let us thus briefly explain why (T5)
still holds.

Assume the context of axiom (T5). Denote the configuration spaces on
$\Sigma$, $\Sigma_1$ and $\Sigma_2$ by $K$, $K_1$ and $K_2$
respectively. Denote the configuration space on
$\Sigma_1\cup\Sigma_2$ by $K_{1 2}$. Also, denote the configuration
spaces on $\Sigma\cup\Sigma_1$ and on $\Sigma\cup\Sigma_2$ by $K_{1s}$ and
$K_{2s}$ respectively.
Furthermore, we denote the corners forming the intersection of
$\Sigma_1$ and $\Sigma_2$ by $c$.
Then, by gluing properties of the path integral we have
\begin{equation}
 Z_M(\varphi_{1 2})=
 \int_{K, \varphi|_c=\varphi_{1 2}|_c} \xD\varphi\,
  Z_{M_1}(\varphi_{1 2}|_{\Sigma_1}\cup\varphi) 
  Z_{M_2}(\varphi_{1 2}|_{\Sigma_2}\cup\varphi) ,
\end{equation}
where $\varphi_{1 2}\in K_{1 2}$ and $\cup$ denotes the joining of
configuration data. Note the formal property of the
ON-basis $\{\xi_i\}_{i\in I}$ of $\cH_\Sigma$,
\begin{equation}
 \sum_{i\in I} \xi_i(\varphi_a)\xi_i^*(\varphi_b)
 =\delta(\varphi_a,\varphi_b)\quad \forall\varphi_a,\varphi_b\in K.
\label{eq:deltaid}
\end{equation}
Here $\xi_i^*$ denotes the dual basis element of $\xi_i$ in
$\cH_{\bar{\Sigma}}$.
Combining those properties with the identity
\begin{equation}
\begin{split}
 \int_{K_{1s}}\xD\varphi_{1s}\,\int_{K_{2s}}\xD\varphi_{2s}\,
 \delta(\varphi_{1s}|_\Sigma,\varphi_{2s}|_\Sigma)
 f(\varphi_{1s},\varphi_{2s})\\
 =\int_{K_{1 2}}\xD\varphi_{1 2}\,
 \int_{K, \varphi|_c=\varphi_{1 2}|_c}\xD\varphi\,
 f(\varphi_{1 2}|_{\Sigma_1}\cup\varphi,
  \varphi_{1 2}|_{\Sigma_2}\cup\varphi)
\end{split}
\end{equation}
for arbitrary $f:K_{1s}\times K_{2s}\to\C$
yields the formula of axiom (T5) as required.

\section{Two-dimensional TQFT with corners}
\label{sec:2dax}

We now specialize to the case of two dimensions. For the moment we
remain in the setting of topological manifolds and also restrict them
to be compact.

Hence, a connected component of a hypersurface is either an oriented
closed interval or an oriented circle. For simplicity, we refer to
the former object as an \emph{open string} and to the latter as a
\emph{closed string} (not mentioning the orientation
explicitly). These are the only \emph{elementary hypersurfaces}. A
general hypersurface is then simply a finite disjoint union of open
and closed strings.

Let us denote the Hilbert spaces associated to the open and closed
string by $\cHo$ and $\cHc$
respectively. We denote their inner products by $\langle
\cdot,\cdot\rangle_\textrm{O}$ and
$\langle\cdot,\cdot\rangle_\textrm{C}$ respectively. We denote the
involutions of (T1b) by $\iotao:\cHo\to\cHob$ and
$\iotac:\cHo\to\cHcb$ respectively. The bar indicates that we are
considering the strings with opposite orientations.

Let us consider the concept of decomposition of a
hypersurface. Axiom (T2) tells us that associated with a decomposition
is a map $\tau$. The composition of the maps $\tau$ of
consecutive decompositions of a hypersurface is the same as the map
$\tau$ associated to the resulting decomposition. Hence, we only need
to consider decompositions with at most two components. What is more,
in the case of disjoint decompositions the map $\tau$ is simply an
identification of the state space with the tensor product of the state
spaces of the components. Thus, it is enough to specify $\tau$ for
decompositions of connected hypersurfaces.

Indeed, it is easy to see that there are merely two \emph{elementary
decompositions}. The first one is the decomposition of an open string
into two open strings. In that case there are two corners at
the two ends of the string to be decomposed and one corner at the
point where the string is to be cut. The latter appears in
both of the component strings and is a virtual corner with respect to
the original string. We denote the induced map by
$\tauo:\cHo\tens\cHo\to\cHo$. Note that the decomposition property
implies that $\tauo$ is \emph{associative}, making $\cHo$ into an
associative algebra.

The other elementary decomposition is that of a closed string into an
open string. In this case we mark one point of the closed string as a
virtual corner and cut it open there. The resulting open string has
two copies of this corner as its endpoints. We denote the induced map
by $\tauc:\cHo\to\cHc$. Note that the composition $\tauc\circ\tauo$
must be \emph{commutative} due to the lack of natural ordering of the two
pieces in the process of cutting a closed string into two pieces.

Compact connected orientable manifolds with boundary are Riemann
surfaces with
holes. Thus, a region is simply a finite union of oriented Riemann
surfaces with holes. A connected region is then characterized by two
non-negative integers, the genus $g$ and the hole number $n$. In view
of the
gluing axiom for regions (T4), however, and taking into account the
fact that virtual corners allow us to perform rather arbitrary
gluings, there is only one \emph{elementary region}. This is the
disc. All other regions can be obtained by gluing discs
together. We denote the associated amplitude by $\rho_D:\cHc\to\C$.

\begin{figure}[h]
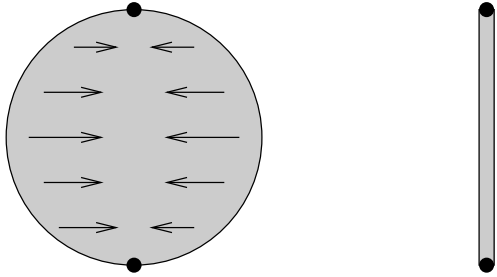

\begin{center}
\begin{tabular}{cp{2cm}c}
\input{fig_disccomp1.pstex_t} &&
\input{fig_disccomp2.pstex_t}
\end{tabular}
\end{center}
\caption{Empty disc and boundary: The boundary of a disc is decomposed
  into two open strings and the disc is ``squeezed'' (left) until
  the two boundary components coincide (right).}
\label{fig:disccomp}
\end{figure}

In the present topological setting a disc is the same as an empty
disc. Hence, we can use
axiom (T3x) to relate the inner product on the open string state space
to the disc amplitude. Namely, we insert two virtual corners on the closed
string boundary of the disc and decompose the boundary into two open
strings which are then identified up to orientation (squeezing the
interior of the disc), see Figure~\ref{fig:disccomp}. Axiom (T3x) then
implies, 
\begin{equation}
 \langle \psi,\eta \rangle_{\textrm{O}} =
 \rho_{D}\circ\tauc\circ\tauo(\iotao(\psi)\tens \eta)\qquad\forall
 \psi,\eta\in\cHo .
\label{eq:edid}
\end{equation}
Note a subtlety here: Formally, the domain of $\tauo$ is
$\cHo\tens\cHo$. However, when shrinking the disc we view the two open
string components of the boundary as oppositely oriented and hence
denote this domain
by $\cHob\tens\cHo$ (composing with $\iotao$ then recovers a domain
$\cHo\tens\cHo$). This apparent ambiguity between $\cHo$ and $\cHob$
is merely a shortcoming of our
notation. To identify a hypersurface as oppositely
oriented to an identical copy makes only sense when the two copies are
geometrically identified, i.e., ``occupy the same space''.

The inner product on $\cHc$ is completely determined by
that on $\cHo$ in combination with the map $\tauc$, see axiom
(T2). Alternatively, the inner product of $\cHc$ is related via (T3x)
to the amplitude of an empty cylinder. The latter can be obtained in
turn by gluing the empty disc to itself in a suitable way.

\begin{figure}[h]
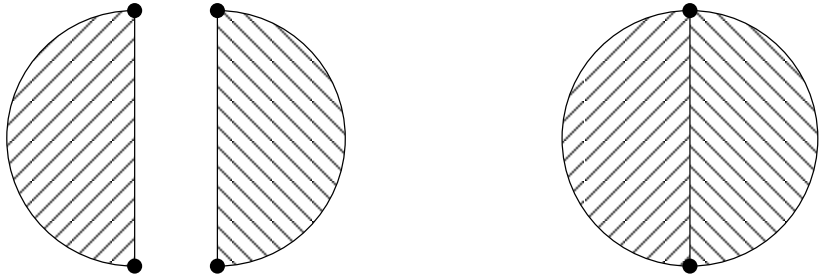

\begin{center}
\begin{tabular}{cp{2cm}c}
\input{fig_discglue1.pstex_t} &&
\input{fig_discglue2.pstex_t}
\end{tabular}
\end{center}
\caption{Gluing two discs to one: Each disc boundary is decomposed
  into two open strings (left), then the discs are glued by
  identifying one of the open strings from each (right).}
\label{fig:discglue}
\end{figure}

In the present topological setting axiom (T4b) is automatically
satisfied for the disc with its boundary decomposed into two open
strings. Indeed, this follows directly from axiom (T3b) discussed
above, using again that the disc is the same as the empty
disc. Similarly, axiom (T5) is automatic for gluing two discs to a new
disc. This involves again decomposing the boundaries into two open
strings, see Figure~\ref{fig:discglue}. Inserting an ON-basis times
its dual into the pair of
extended disc amplitudes yields again the extended disc
amplitudes. This is obvious from interpreting the extended amplitude
as the bilinear from of (T3x). Then, everything descends from extended
amplitudes to non-extended amplitudes, yielding (T5).

We have identified elementary data that completely determine a theory
satisfying the axioms in the case of 2-dimensional compact
topological manifolds. However, this data is not free, but
subject to several conditions, some of which we have identified.

\section{Two-dimensional quantum Yang-Mills theory}
\label{sec:2dqym}

Two-dimensional quantum Yang-Mills theory on arbitrary compact
surfaces was solved in the early 1990s
\cite{Wit:qgauge2d,Fin:qymriem,Rus:gauge2d,BlTh:qym}. It is only a
small step from there to an explicitly TQFT-like
formulation. Our main interest here, however, is the additional step
to extend this to the case with corners.
As we shall see, this provides a realization of the
axioms introduced in Section~\ref{sec:axiomlist}.
To expose the novel aspects of our treatment in detail we proceed in
an essentially self-contained fashion.

\subsection{Path integral and gauge symmetry}

Let $G$ be a compact, connected and simply connected Lie group.
The field of classical Yang-Mills theory is given by a
connection 1-form $A$ for a principal $G$-bundle on a manifold $M$
with metric. (We restrict to the case of the principal bundle being
trivial.)
The Yang-Mills
action is
\begin{equation}
 S_M[A]=-\frac{1}{\gamma^2}\int_M \tr(F\wedge\star F) ,
\label{eq:act}
\end{equation}
where $F$ is the curvature 2-form of $A$ and $\gamma$ the (classically
irrelevant) coupling constant.
If $M$ is two-dimensional,
the dependence of $S_M[A]$ on the geometry of $M$ is merely through  
the area form of the metric.\footnote{Note that this applies whether the
metric is Riemannian or pseudo-Riemannian.}

It turns out that the naive version of the Schr\"odinger-Feynman
approach to quantization of Section~\ref{sec:quantize} is not quite
appropriate as we have to take into account gauge symmetry.
We shall see how this can be accomplished through a suitable
modification of the procedure.

Consider a manifold $M$ with the topology of a disk. Its boundary
$\partial M$ is a closed string and we denote a field configuration on
it by $A_\partial$. Naively implementing a Schr\"odinger
representation, $A_\partial$ is simply a connection 1-form on
$\partial M$. Formally, the
field propagator (\ref{eq:prop}) is thus the path integral
\begin{equation}
 Z_M[A_\partial]=\int_{A|_{\partial M}=A_\partial} \xD A\,
 e^{\im S_M[A]} .
\label{eq:ymprop}
\end{equation}
Here, the integral is over connection 1-forms $A$ in $M$ which
restrict on the boundary to $A_\partial$. Note that the argument of
the exponential is imaginary as it should be in quantum
theory. Indeed, this will be essential for unitarity and the
probability interpretation.\footnote{This has nothing to do here with
with the choice of the signature of the metric.}

By gauge invariance $Z_M$ can only depend on
the holonomy $g$ of the boundary connection $A_\partial$ around the
boundary. What is more, $Z_M$ can only depend on the conjugacy class
of $g$ (which also makes the choice of the starting point for the
holonomy irrelevant).
On the other hand, $Z_M$ should be invariant under
orientation preserving diffeomorphisms of $M$ that map the boundary to
itself.
Note that the restriction of such a diffeomorphism to the
boundary leaves the conjugacy class of the holonomy $g$ of the boundary
connection $A_\partial$ invariant.
Hence, the only geometric
information relevant to the value of $Z_M$ can be in the total area of
$M$.\footnote{A diffeomorphism can transform a disc with an
  area form to any other disc with a given area form if and only if the
  total areas are equal.}

Obviously, the value of $Z_M$ should not change whether we
introduce a corner (in the differentiable sense) in its boundary or
smooth it off, preserving the area. Hence, the differentiable
structure of $M$ is expendable as well. This leaves us in an almost
topological setting.
Thus, a region is a pair
$(M,s)$ of a two-dimensional compact oriented topological manifold
$M$ with boundary and a non-negative real number $s$, the area. 
If $s=0$ the region is an empty region.
If we glue two regions $(M,s)$ and $(N,t)$, the area of the new region
is the sum of the areas, i.e., we get $(M\cup N,s+t)$.
A hypersurface
does not carry any additional structure. Hence it is simply a
one-dimensional compact oriented topological manifold with
boundary.

\subsection{Hypersurfaces and state spaces}

The above analysis of gauge symmetry tells us that the appropriate
reduced configuration space for the closed string is the space of
conjugacy classes of $G$. Thus, the state space $\cHc$
should be the space of functions on it. Before returning to this
space, let us consider more general functions on $G$.
The group $G$ has a unique normalized and
invariant measure, the Haar measure. This yields the inner product
\begin{equation}
 \langle \psi, \eta\rangle = \int \xd h\, \overline{\psi(h)}\eta(h)
\label{eq:ipgroup}
\end{equation}
for functions $\psi$ and $\eta$. The set of square integrable
functions $\falg(G)$ on $G$ becomes a Hilbert space with this inner
product. We denote by $\falgc(G)$ the closed subspace of class
functions, i.e.,
functions $\psi$ that are invariant under conjugation, $\psi(g)=\psi(h
g h^{-1})$ for all $g,h\in G$. Hence, this subspace can be thought of
as a space of functions on the conjugacy classes of $G$. This provides
$\cHc$ and its inner product.

What is the state space for an open string? One way to think about
this is to consider the above example of the disc propagator, but
think of the boundary as decomposed into several open
strings. Obviously, if we know the holonomy along each open string
we can calculate the total holonomy
and that is all we need. Hence, it is sufficient to associate a group
element with each open string representing this holonomy.
There are gauge transformations that change the values of these
group elements. However, they necessarily change several group
elements simultaneously (except if there is only a single one).
In considering a single open string alone such
gauge transformations cannot be permitted as we do not know about
other strings we might want to attach to it.
In other words, for determining the
configuration space associated to an open string only gauge
transformations are relevant that act identical on the ends of the
string. Thus, the associated configuration space is the space of
elements of $G$ and the state space is the space of functions on
it. Summarizing, we get
\[
 \cHo=\falg(G),\qquad \cHc=\falgc(G),
\]
with the inner product given in both cases by (\ref{eq:ipgroup}).
We thus have specified the realization of axiom (T1) for elementary
(in the sense of Section~\ref{sec:2dax}) hypersurfaces.

A suitable orthogonal basis for $\falg(G)$ is given through the
Peter-Weyl
decomposition by matrix elements of irreducible representations. We
denote these by $t^V_{i j}$, where $V$ is the representation and $j$
and $i$ are indices for a basis of $V$ and its dual respectively. A
suitable orthogonal basis for the subspace $\falgc(G)$ is given by the
characters associated with irreducible representations, $\chi^V=\sum_i
t^V_{i i}$. Recall the identities,
\begin{gather*}
\overline{t^V_{i j}(g)}=t^V_{j i}(g^{-1})\quad\text{and}\quad
\int\xd g\, t^V_{i j}(g^{-1}) t^W_{m n}(g)
=\delta_{V,W}\delta_{i,n}\delta_{j,m}\frac{1}{\dim V} .
\end{gather*}
Hence, the inner product (\ref{eq:ipgroup}) with respect to these basis
elements is
\begin{equation}
 \langle t^V_{i j},t^W_{m n}\rangle=
 \delta_{V,W}\delta_{i,m}\delta_{j,n}\frac{1}{\dim V},\quad
 \langle \chi^V,\chi^W\rangle=
 \delta_{V,W} .
\label{eq:ipm}
\end{equation}

According to (\ref{eq:conj}) we might expect the antilinear
involution of axiom (T1b) to be simply complex conjugation of the
wave function. However, this is not the case here because the
configuration data is sensitive to the orientation of a
hypersurface. More, precisely, a holonomy $g$ along an open string
(for example) becomes a holonomy $g^{-1}$ if the look at the string
``the other way round'', i.e.\ change its orientation. This was not
taken into account in Section~\ref{sec:quantize}. Hence, the
antilinear involution is really given by
\begin{equation}
 (\iotao(\psi))(g)=\overline{\psi(g^{-1})},\qquad
 (\iotac(\psi))(g)=\overline{\psi(g^{-1})} .
\end{equation}
In terms of matrix elements this is
\begin{equation}
 \iotao(t^V_{i j})=t^V_{j i},\qquad
 \iotac(\chi^V)=\chi^V . 
\label{eq:iotam}
\end{equation}

We now turn to axiom (T2) describing hypersurface decompositions. As
explained in Section~\ref{sec:2dax} there are only two elementary
ones. We consider the decomposition of an open string into two open
strings first. Gauge symmetry means that we do not actually have a map
$G\to G\times G$ that expresses the splitting of configuration
data. Indeed there are many ways a holonomy $g$ could be split into a
product $g_1 g_2$. However, since we are really dealing with functions
on configuration data we can solve this problem by integrating over
all such splittings. Hence,
\begin{equation}
 \left(\tauo(\psi\tens\eta)\right)(g)
 = \int\xd h\, \psi(g h)\eta(h^{-1})
 = \int\xd h\, \psi(h)\eta(h^{-1} g).
\label{eq:tauo}
\end{equation}
Note that this looks quite different from (\ref{eq:tauphi}).
Indeed, the integral appearing in
$\tauo$ may be seen as an averaging over gauges. These are precisely
the ``missing'' gauge transformations at the endpoints of open strings
that we can only perform once we attach the open string to something
else. Note also the analogy to gauge transformations in lattice gauge
theory. These are performed at a vertex and affect the holonomies
associated with all edges connected to the vertex.
In terms of matrix elements,
\begin{equation}
 \tauo(t^V_{i j}\tens t^W_{m n})
 =\delta_{V,W}\delta_{j,m}\frac{1}{\dim V}t^V_{i n} .
\end{equation}
This map is indeed associative as required for consistency (see
Section~\ref{sec:2dax}). It makes $\cHo$ into an associative
algebra which is commutative only if $G$ is abelian.
Note that the algebra product is quite different from the commutative
algebra product of $\cHo$ as an algebra of functions.

The only other elementary hypersurface decomposition is that of a
closed string into an open string. On the level of configuration data
we want to recover a group element from its conjugacy class. Again,
(in the non-abelian case) there are many possible group elements which
yield the same conjugacy class. Hence, we integrate over them,
\begin{equation}
 \left(\tauc(\psi)\right)(g)=\int\xd h\, \psi(h g h^{-1}).
\label{eq:tauc}
\end{equation}
Again, this may be seen as an averaging over gauges at the endpoint where
we glue the open string to itself. In terms of matrix elements,
\begin{equation}
 \tauc(t^V_{i j})
 =\delta_{i,j}\frac{1}{\dim V}\chi^V .
\end{equation}
Indeed, this is an orthogonal projection operator from the Hilbert
space $\cHo$ to its subspace $\cHc$. If $G$ is abelian, $\cHo=\cHc$
and $\tauc$ is simply the identity.
The composition
\begin{equation}
\tauc\circ\tauo(t^V_{i j}\tens t^W_{m n})
 =\delta_{V,W}\delta_{j,m}\delta_{i,n}\frac{1}{(\dim V)^2}\chi^V
\label{eq:dec2}
\end{equation}
is commutative as required for consistency (see
Section~\ref{sec:2dax}). It is also straightforward to check axiom
(T2b) explicitly.

\subsection{The disc region and amplitudes}

Let us return to the propagator (\ref{eq:ymprop}). As we have seen
this depends only on the conjugacy class of the holonomy $g$ around the
boundary. Hence we, can expand it in characters. The dependence on the
geometry of the disc manifold $M$ is only through its area
$s$. Therefore, the expansion coefficients depend only on this real
number $s$ and we can write
\begin{equation}
 Z_M[g]=\sum_V \dim V\, \alpha_V(s)\, \chi^V(g) ,
\label{eq:discprop}
\end{equation}
where the sum is over the finite-dimensional irreducible
representations $V$ of $G$.
Without knowing the exact nature of the functions $\alpha_V(s)$ we can
write the amplitude map (\ref{eq:ampl}) for the disc $D$ with area $s$ as
\begin{equation}
 \rho_{(D,s)}:\cHo\to\C\qquad \rho_{(D,s)}(\psi)
 =\sum_V \dim V\, \alpha_V(s)\,\int\xd g\,  \chi^V(g) \psi(g) .
\end{equation}
In terms of matrix elements this is simply
\begin{equation}
\rho_{(D,s)}(\chi^V)
 =\dim V\, \alpha_V(s) .
\end{equation}

Recall from Section~\ref{sec:2dax} that the disc is the only elementary
region out of which we can construct any other region by
gluing. Hence, we need not specify any other amplitude a
priori. However, we have several consistency conditions. In
particular, recall from Section~\ref{sec:2dax} that axiom (T3x)
implies the identity (\ref{eq:edid}). We start by considering the
extended amplitude map
$\rho_{(D,s)}\circ\tauc\circ\tauo$ arising from decomposing the
boundary of a disc into two open strings. Composing (\ref{eq:dec2})
with the amplitude yields
\begin{equation}
 \rho_{(D,s)}\circ\tauc\circ\tauo(t^V_{i j}\tens t^W_{m n})
 =\delta_{V,W}\delta_{j,m}\delta_{i,n}\frac{\alpha_V(s)}{\dim V} .
\label{eq:edampl}
\end{equation}
When shrinking the disc (recall Figure~\ref{fig:disccomp}), the two
open strings
coincide up to orientation and the above map (with $s=0$) is
interpreted as a pairing
$(\cdot,\cdot)_\textrm{O}:\cHob\tens\cHo\to\C$.  
Using (\ref{eq:ipm}) and (\ref{eq:iotam})
the requirement $\langle\cdot,\cdot\rangle_\textrm{O}
=(\iotao(\cdot),\cdot)_\textrm{O}$ is then seen to be equivalent to the
condition that $\alpha_V(0)=1$ for all irreducible representations
$V$. Thus, the pairing is explicitly given by
\begin{equation}
 (t^V_{i j}, t^W_{m n})_\textrm{O}
 =\delta_{V,W}\delta_{j,m}\delta_{i,n}
 \frac{1}{\dim V} .
\end{equation}
In particular, taking $\{t^V_{i j}\}_{V,i,j}$ as a basis of $\cHo$,
the dual basis of $\cHob$ is given by
$\{\dim V\, t^V_{j i}\}_{V,i,j}$. This satisfies 
equation (\ref{eq:deltaid}), modified to take into account the
orientation dependence of the holonomies:
\begin{equation}
 \sum_{V,i,j} \dim V\, t^V_{i j}(g) t^V_{j i}(h)=\delta(g,h^{-1}) .
\end{equation}

Without working
out the amplitude for the empty cylinder at this point, we know from
(\ref{eq:ipm}) and (\ref{eq:iotam}) that the pairing
$(\cdot,\cdot)_\textrm{C}:\cHcb\tens\cHc\to\C$ must be given by
\begin{equation}
 (\chi^V, \chi^W)_\textrm{C}
 =\delta_{V,W} .
\end{equation}
Hence, taking $\{\chi^V\}_V$ as a basis of $\cHc$ the dual basis of
$\cHcb$ is given by $\{\chi^V\}_V$.

Consistency requires that gluing the disc to another
disc satisfies axiom (T5). This is no longer automatic as in the
purely topological context of Section~\ref{sec:2dax}.
Rather, ``composing'' the amplitude
for a disc with area $s_1$ with the amplitude of a disc with area $s_2$
should yield the amplitude for a disc with area
$s_1+s_2$. Geometrically, we need to decompose the boundary of each of
the two discs to be glued into two open strings. One open string of
one disc is then glued to an oppositely oriented open string of the
other disc, recall Figure~\ref{fig:discglue}.
Algebraically, this yields the identity,
\begin{multline}
 \rho_{(D,s_1+s_2)}\circ\tauc\circ\tauo(\psi\tens\eta)\\
 =\sum_{V,i,j}\dim V\,\rho_{(D,s_1)}
 \circ\tauc\circ\tauo(\psi\tens t^V_{i j})
 \rho_{(D,s_2)}\circ\tauc\circ\tauo(t^V_{j i}\tens\eta) .
\end{multline}
Note that we have used the basis $\cHo$ and dual basis of $\cHob$ as
determined above.
Evaluating this on matrix elements using (\ref{eq:edampl}) yields the
series of identities 
$\alpha_V(s_1+s_2)=\alpha_V(s_1)\alpha_V(s_2)$ for all $V$. Together
with the condition $\alpha_V(0)=1$ derived above and assuming
continuity of the functions $\alpha_V$ we find that they must be
exponentials of the area $s$. That is,
$\alpha_V(s)=\exp(\beta_V\,s)$ for unknown constants $\beta_V$.

The only axiom we have not used so far is the unitarity axiom
(T4b). This stands apart as it is much more related to the physical
interpretation of the formalism than to its mathematical coherence.
The simplest context for its application is given by decomposing the
boundary of a disc into two open strings and converting the extended
amplitude map into a map between the state spaces associated with the
open strings. Concretely, we have to dualize one tensor component in
the domain of the map (\ref{eq:edampl}).
This yields a linear
map $\tilde{\rho}_{(D,s)}:\cHo\to\cHo$ given by
\begin{equation}
 \tilde{\rho}_{(D,s)}(t^V_{i j})=\exp(\beta_V\,s)\, t^V_{i j} .
\label{eq:dtampl}
\end{equation}
This defines obviously an isomorphism on the vector space of matrix
elements which is dense in $\cHo$. We declare
that it should in fact be a vector space isomorphism on the whole
space $\cHo$. Then axiom (T4b) requires unitarity. This means the
condition $|\exp(\beta_V\,s)|=1$ for any irreducible
representation $V$ and any area $s$. If we equipped
all regions with zero area this would be automatically
satisfied. Indeed, this would make the theory topological and we have
already seen in Section~\ref{sec:2dax} how this implies unitarity. In
general, however, we find that the constants $\beta_V$ must be
imaginary to satisfy this condition.

\subsection{General regions}

As already mentioned, the amplitude for any region can be
obtained via gluing from that of the disc. What is more, the
amplitude for any connected region can be obtained by taking a single
disc, decomposing its boundary suitably and then gluing pieces of this
boundary together in suitable ways. Gluing a region to itself is not
explicitly mentioned in axiom (T5), but it is implicitly obtained by
gluing with an empty region.

We consider explicitly here only the cylinder. To obtain it, we
decompose the boundary of a disc into four open strings and glue two
non-adjacent ones together.
\begin{multline}
 \rho_{(\textrm{cyl},s)}\circ(\tauc\tens\tauc)(\psi\tens\eta)\\
 =\sum_{V,i,j} \dim V\,\rho_{(D,s)}
  \circ\tauc\circ\tauo\circ(\tauo\tens\tauo)
  (\psi\tens t^V_{i j}\tens\eta\tens t^V_{j i}) .
\end{multline}
Concretely, $\rho_{(\textrm{cyl},s)}$ can be expressed as
\begin{equation}
 \rho_{(\textrm{cyl},s)}(\psi\tens\eta)
 =\sum_V \dim V\, \exp(\beta_V\, s)\,
  \int\xd g\,\xd h\,  \chi^V(g) \psi(g h^{-1}) \eta(h) ,
\label{eq:cylampl}
\end{equation}
or in terms of characters,
\begin{equation}
 \rho_{(\textrm{cyl},s)}(\chi^V\tens \chi^W)
 =\delta_{V,W} \exp(\beta_V\, s) .
\end{equation}
At this point, we may easily verify axiom (T3x) for the empty
cylinder.

Any connected region is a Riemann surface with holes. We classify it
by genus $g$ and hole number $n$. To obtain the amplitude for
any Riemann surface we may for example first work out the amplitude
for a sphere with $n+2g$ holes and then glue $g$ pairs of holes
together. The result is,
\begin{equation}
  \rho_\textrm{g,n,s}(\chi^{V_1}\tens\cdots\tens\chi^{V_n})
 =\delta_{V_1,\dots,V_n}\exp(\beta_{V_1}\, s)\,(\dim V_1)^{2-2 g -n} .
\end{equation}
In the special case with no hole we obtain also a sum over
representations,
\begin{equation}
  \rho_\textrm{g,0,s}
 =\sum_V\exp(\beta_{V}\,s)\,(\dim V)^{2-2 g} .
\label{eq:amplcl}
\end{equation}
These amplitudes reproduce well known formulas (obtained without
corners), see \cite{Wit:qgauge2d,BlTh:qym}.

\subsection{The constants $\beta_V$}

So far we have not really exploited the actual form of the Yang-Mills
action. We have only used (a) the fact that the action
depends only on an area form, (b) gauge symmetry, 
(c) general properties of the path integral
(\ref{eq:ymprop}), and (d) the unitarity
requirement. This has allowed us to determine the theory completely up
to a set of unknown imaginary numbers $\beta_V$, one for each irreducible
representation $V$ of $G$.

A more detailed analysis of the path integral (\ref{eq:ymprop})
shows that $\beta_V$ takes the form
\begin{equation}
\beta_V=\frac{\im}{4} \gamma^2 C_V,
\label{eq:betav}
\end{equation}
where $C_V$ is the value of the quadratic Casimir operator on the
representation $V$. This is well known in lattice gauge theory, see
e.g.\ \cite{Mig:reclgt}. For a
simple derivation using essentially the same conventions as
here (except for the $\im$), see \cite{Oe:tqft}.

Unsurprisingly, exchanging the $\im$ in the path integral
(\ref{eq:ymprop}) for a $-1$ (as customary for example in lattice
gauge theory) effects the same change in $\beta_V$, making it
real. What is more, this makes the sums over irreducible
representations appearing above generally convergent. While we have
not mentioned this explicitly so far, those sums are not guaranteed to
converge in our setting.
However, amplitudes generally do converge if the boundary states
are matrix elements. For example,
we really have defined the amplitude map for the disc
only on the dense domain of the Hilbert space spanned by
matrix elements. Indeed, one can check that with $\beta_V$ defined
by (\ref{eq:betav}) this amplitude map is unbounded and hence cannot
be continuously extended to the whole Hilbert space. However, we may
still formally satisfy axiom (T4) if we find a non-continuous
extension to the whole Hilbert space. The physical relevance of this
is questionable though and we do not pursue this point further.

Other amplitudes that may be ill-defined are those for closed regions
given by (\ref{eq:amplcl}). Notably, the amplitudes for the sphere
($g=0$) and the torus ($g=1$) will be ill-defined. The behaviors of
amplitudes for higher genus surfaces depend on the group. For an
abelian group all these amplitudes will diverge while for example for
$SU(2)$ the ones for $g\ge 2$ will converge. Most of these problems can
be avoided if we make $\beta_V$ real. However, in the case of zero
area $s=0$, i.e., for empty regions, they will persist.

\subsection{Unitarity and probability conservation}

The main reason for insisting on imaginary $\beta_V$ is the
physical interpretation. In quantum mechanics, unitarity is essential
for a sensible probability interpretation. However, this is usually
thought to make sense only for transitions between spacelike
hypersurfaces. In the present context, we could for example take a
cylindrical spacetime with space being the circle and time an
interval. In this example axiom (T4b) has indeed the usual meaning of
unitarity as guaranteeing probability conservation.

It was shown in \cite{Oe:GBQFT} that a more general
probability interpretation is possible which is not restricted to
spacelike hypersurfaces. (For an actual example with timelike
hypersurfaces see \cite{Oe:KGtl}.) The unitarity axiom (T4b) then
acquires the physical meaning of probability conservation in more general
situations.

\begin{figure}[h]
\begin{center}
\input{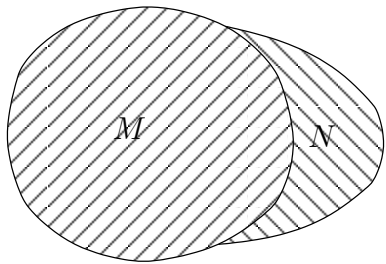}
\end{center}
\caption{A disc $M$ with an adjacent smaller disc $N$,
  deforming it.}
\label{fig:deform1}
\end{figure}

In particular, consider a region $M$ which we take to be a disc
here. Now, enlarge the disc $M$ through a ``small'' outward
deformation $N$ to a new region $M\cup N$. For simplicity we let $N$
and $M\cup N$ be (topological) discs as well, see
Figure~\ref{fig:deform1}. Denote the state
spaces associated with the boundaries of $M$ and $M\cup N$ by $\cH_M$
and $\cH_{M\cup N}$. Probability is then conserved for measurements
associated to the boundary of $M$ relative to measurements associated
to the boundary of $M\cup N$ if the map $\cH_{M\cup N}\to\cH_M$
induced by the amplitudes is unitary. As was already remarked in
\cite{Oe:GBQFT} this setup necessarily involves corners. Indeed, this
provides a major reason for our interest in corners here.

\begin{figure}[h]
\begin{center}
\input{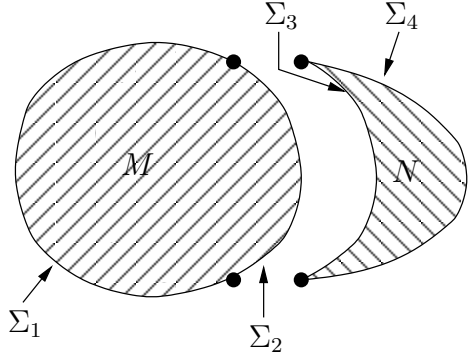}
\end{center}
\caption{Decomposition of the boundaries of $M$ and $N$ into open
  string hypersurfaces.}
\label{fig:deform2}
\end{figure}

It is easy here to compute the map in question. Decompose the
boundaries of $M$ and $N$ into two open strings each, so that they
have one open string forming their common boundary. This is shown in
Figure~\ref{fig:deform2}. Now, consider the map
$\tilde{\rho}_N:\cH_{\Sigma_4}\to\cH_{\Sigma_3}$ induced by the
extended amplitude of
the disc $N$. As we have seen this is given by (\ref{eq:dtampl}) with
$s$ the area of $N$.
The induced map from the decomposed
boundary state space $\cH_{\Sigma_1}\tens\cH_{\Sigma_4}$ of $M\cup N$
to the decomposed boundary state space
$\cH_{\Sigma_1}\tens\cH_{\Sigma_2}$ of $M$ is simply the product
$\id_{\Sigma_1}\tens\tilde{\rho}_N$. (Recall that $\Sigma_3$ and
$\Sigma_2$ are
identified.) It remains to obtain the corresponding map
$\cH_{M\cup N}\to\cH_M$ between the undecomposed boundary state
spaces. This amounts to completing the bottom line of the commutative
diagram,
\[
\begin{CD}
\cH_{\Sigma_1}\tens\cH_{\Sigma_4} @>>>
\cH_{\Sigma_1}\tens\cH_{\Sigma_2} \\
@VVV @VVV \\
\cH_{M\cup N} @>>> \cH_M
\end{CD}
\]
where the vertical maps are given by (\ref{eq:dec2}).
It is easy to see that the required map is given by
$\chi_V\mapsto\exp(\beta_V s)\chi_V$, with $s$ again the area of
$N$. Imaginarity of the $\beta_V$ implies unitarity of this map
and hence probability conservation.

\begin{figure}[h]
\begin{center}
\input{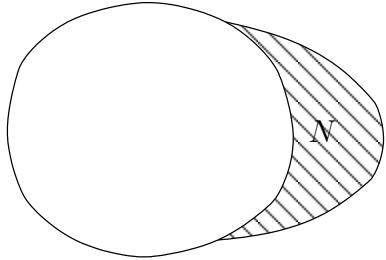}
\end{center}
\caption{The region $N$ as a degenerate cylinder.}
\label{fig:deform3}
\end{figure}

The above result is exactly the map one obtains from converting the
cylinder amplitude (\ref{eq:cylampl}) into a map between its two
bounding closed
string state spaces. This is not surprising. Indeed, we could
avoid the use of corners above by thinking of $N$ as a cylindrical
region that surrounds $M$ completely. This cylinder would be
infinitely thin along parts of its boundaries, representing a kind of
``partially empty'' region, see Figure~\ref{fig:deform3}. Our above
computation shows that both pictures are consistent and yield the same
result as was in fact already anticipated in \cite{Oe:GBQFT}. We refer
the reader interested in a more general perspective on these issues to
Sections~4.3 and 9 of that paper.

\section{Extensions}
\label{sec:extensions}

\subsection{The vacuum}

In \cite{Oe:GBQFT} a proposal was made for the axiomatization of the
concept of \emph{vacuum}. This was successfully tested in the context
of Klein-Gordon quantum field theory \cite{Oe:timelike,Oe:KGtl}.
It can be easily adapted and
generalized to the case with corners.

\begin{itemize}
\item[(V1)] For each hypersurface $\Sigma$ there is a distinguished state
  $\psi_{\Sigma,0}\in\cH_\Sigma$, called the \emph{vacuum state}.
\item[(V2)] The vacuum state is compatible with the involution. That is,
  for any hypersurface $\Sigma$,
  $\psi_{\bar{\Sigma},0}=\iota_\Sigma(\psi_{\Sigma,0})$.
\item[(V3)] The vacuum state is compatible with
  decompositions. Suppose the hypersurface
  $\Sigma$ decomposes into components
  $\Sigma_1\cup\dots\cup\Sigma_n$. Then
  $\psi_{\Sigma,0}=\tau(\psi_{\Sigma_1,0}\tens\cdots\tens\psi_{\Sigma_n,0})$.
\item[(V5)] The amplitude of the vacuum state is
  unity, $\rho_M(\psi_{\partial M,0})=1$.
\end{itemize}

Note that axiom (V4) of \cite{Oe:GBQFT} is redundant here as
it is implied by (V5) applied to empty regions. The only other change
is the generalization of axiom (V3). This is
formulated now with the generalized notion of decomposition and
includes the map $\tau$ explicitly.

There is a unique realization of these axioms in two-dimensional
quantum Yang-Mills theory. In both $\cHc$ 
and $\cHo$ the vacuum state is the state $\mathbf{1}$.
This is the constant
function on $G$ with value $1$. In matrix element notation this is
$\mathbf{1}=\chi^0=t^0_{0 0}$, where $0$ denotes the trivial
representation. Note that the quadratic Casimir operator on the
trivial representation is zero, $C_0=0$ and hence $\beta_0=0$ and
$\alpha_0=1$. Verification of the vacuum axioms is elementary.

\subsection{Wilson loops}
\label{sec:wilson}

It is easily possible to implement Wilson loops into the
formalism. The most natural way to do this is via
the introduction of additional labeled empty regions. Since we have
two types of elementary hypersurfaces, there are correspondingly two
types of elementary labeled regions. A labeled empty region
with the shape of an open string is labeled by an element of
$\cHo$. We denote the corresponding extended amplitude map
$\cHo\tens\cHob\to\C$ by $\rho_{\textrm{O},\psi}$, where the label is
$\psi\in\cHo$. It is defined via
\begin{equation}
 \rho_{\textrm{O},\psi}(\eta\tens\mu)
 =\int\xd g\, \eta(g)\psi(g)\mu(g^{-1}) .
\label{eq:wilsono}
\end{equation}
Note that the orientation enters here in a special way. We can think
of $\psi$ as associated with an oriented piece of loop that points in
the same direction as the orientation of the ``side'' which carries
$\eta$, but oppositely to the one that carries $\mu$. This is reflected
in how $g$ enters in the arguments of the different functions either
as $g$ or as $g^{-1}$.

Similarly, we get labeled empty regions with the shape of a closed
string. The label set is now the state space $\cHc$. We denote the
corresponding extended amplitude map $\cHc\tens\cHcb\to\C$ by
$\rho_{\textrm{C},\psi}$, where the label is $\psi\in\cHc$. It is
defined with the same formula as (\ref{eq:wilsono}), except that all
functions must now be class functions.

Note that the newly defined empty regions generalize the empty disc
and the empty cylinder. Indeed, these are recovered for the special
choice of label $\psi=\mathbf{1}$, the constant function with value
$1$. However, the newly defined empty regions have in general no
version of the amplitude with ``non-decomposed'' boundary as the empty
disc and cylinder have.

On first sight it may seem that our definitions have little
resemblance to what one usually considers as Wilson loops. However,
the new empty regions are precisely closed Wilson loops (closed
string) or pieces thereof (open string). In the closed string case the
label is a class function. In particular, we can choose a
character. This recovers the usual labeling of Wilson loops by
irreducible representations.

In the
open case we have only a piece of a Wilson loop. To obtain a closed
Wilson loop we have to glue pieces together. This gluing is not
directly a gluing of the new empty manifolds at their endpoints as
this notion does not exist. However, we may deduce such a notion by
making more complicated gluings involving empty discs.
The result is
that the gluing of two labeled empty open string regions to a
labeled empty open string region is given by formula
(\ref{eq:tauo}). Similarly, the gluing of a labeled empty open
string region to itself by joining the endpoints is given by formula
(\ref{eq:tauc}). Hence, we obtain a new and completely different
interpretation of the maps $\tauo$ and $\tauc$. Instead of applying
them to states we apply them to labels here. In particular, we see
that as soon as we close a Wilson loop we get a label by a class
function, or if we use matrix elements, by an irreducible
representation.

It is relatively easy to see that the introduction of the new objects
preserves the consistency and coherence of the axioms. Since our
interest here is merely the implementation in principle,
we abstain from
performing concrete but
straightforward calculations of amplitudes for surfaces with inserted
Wilson loops. We refer the reader interested in this to some results
obtained in \cite{Wit:qgauge2d,BlTh:qym} (for closed Wilson loops).

\subsection{Non-orientable surfaces}

So far we have required regions to be orientable. However, this is not
really essential. In axiom (T4) the orientation of the region is used to
induce an orientation on its boundary. Instead, we may drop the
orientation of the region, but explicitly specify an orientation on
the boundary. The latter is essential and implies that regions must
still have orientable boundaries. In axiom (T5) the matching
orientations of the regions to be glued are used to ensure that the
boundary components to be glued have opposite orientation. For
non-oriented (including non-orientable) regions we may simply demand
the latter property explicitly.

A problem occurs in so far as the gluing axiom is now less powerful
than it should be. In particular, it will not be possible to obtain a
non-orientable region out of orientable ones through gluing. Thus, in the
two-dimensional case, the disc is no longer elementary in the sense
that every other region can be obtained by gluing discs together. To
remedy this we need to introduce gluings also along hypersurfaces with
parallel instead of opposite orientation. This can be accomplished for
example by inserting an orientation-changing map
$\cH_\Sigma\to\cH_{\bar{\Sigma}}$ into the gluing. Note that this map
must be linear and hence cannot be the antilinear involution $\iota$.

In the context of the two-dimensional quantum Yang-Mills
theory it is completely clear what this map is. Namely, it
corresponds to interpreting the same configuration data on a
hypersurface with opposite orientation. Hence, it corresponds to the
map $g\mapsto g^{-1}$ for holonomy data. This means on the level of
state spaces the map $\psi\mapsto\psi'$ with $\psi'(g)=\psi(g^{-1})$.

We are now in a position to work out amplitudes for non-orientable
surfaces. As a first example consider the M\"obius strip.
To obtain its amplitude we start with a disc and decompose its
boundary into four open strings. We then glue two non-adjacent open
strings together (similarly to the case of the cylinder), but such
that their orientations are parallel. That is, in this gluing we have
to insert the aforementioned map. The resulting
amplitude is
\begin{equation}
 \rho_{(\textrm{M\"ob},s)}(\chi^V)
 =\delta_{V,V^*}\, \exp(\beta_V\, s).
\end{equation}
Here, $V^*$ denotes the representation dual to $V$. In other words,
this amplitude is non-zero only if the representation $V$ is
self-dual. For example, for the group $SU(2)$ this is always the case
while for the group $U(1)$ this is only the case for the trivial
representation.

As another example, the amplitude for the Klein bottle is
\begin{equation}
 \rho_{(\textrm{Klein},s)}=\sum_{V}\delta_{V,V^*}\exp(\beta_V\,s).
\end{equation}
(Glue the cylinder to itself with an orientation reversal inserted.)
The appearance of a factor $\delta_{V,V^*}$ is a general
feature of amplitudes of non-orientable surfaces.

Our method of dealing with non-orientable surfaces
by inserting an orientation changing map is a direct generalization of
the method used by Witten \cite{Wit:qgauge2d} to the case with
corners. We also refer to Witten's article for more examples of
amplitudes for non-orientable surfaces.

\section{Outlook}

Physical experiments are usually confined to finite regions of
spacetime and indpenedent of what goes on in other parts of
spacetime. Hence, it is desirable to be able to describe a physical
process through states and amplitudes associated to such a
region. Furthermore, it is desirable to be able to
describe what happens when we join two such processes and associated
regions together. This is the motivation (in the context of quantum
field theory) for a gluing axiom of the form (T5). However, for
spacetime regions with generic topology like that of a 4-ball
this can only work if we have the concept of corners at our disposal.

At the same time, this gluing with corners yields valuable consistency
conditions. In the example of two-dimensional quantum Yang-Mills
theory, we have seen this explicitly. Gluing the disc to itself to
obtain a new disc yielded strong constraints on the possible form of
the amplitude map. More precisely, the unknown functions $\alpha_V(s)$
could in this way be constrained to be of an exponential form
$\alpha_V(s)=\exp(\beta_V\, s)$.

This same type of gluing could also be made to play a role in the
procedure of
renormalization. For example, we might apply the present formalism to
lattice gauge theory. Instead of the often used toroidally compactified
versions of spacetime, one would consider bounded hypercubic pieces of
spacetime. These would carry a hypercubic lattice within them (and on
their boundary) of given length scale. Comparing the amplitude of one
such piece with the amplitude of several pieces glued together, but
with the lattice scale changed to obtain the same physical dimensions
would allow to set up the corresponding renormalization group
equation. Of course, there are a lot of open questions to be addressed
before this could become viable, such as the identification of
appropriate boundary states.

Spin foam models have recently become popular in approaches to quantum
gravity, see
\cite{Ori:spinfoamrev,Per:sfmodels,Rov:qg,Oe:tqft} and references
therein. These models are state sum
models, but share TQFT-like features in that they are composed out of
elementary building blocks (usually $n$-simplices with certain
labels). It should be possible to bring some of these models into the
axiomatic form of Section~\ref{sec:axiomlist}. This would in turn
allow the application of the probability interpretation proposed in
\cite{Oe:GBQFT} and possibly help resolve long standing problems
regarding their physical interpretation.

Combining this with the relation between gluing and renormalization
suggested above leads to an enhancement of the framework for the
renormalization of spin foam models proposed in
\cite{Oe:renormdisc,Oe:rensfqg}, see also \cite{Oe:tqft}.
We merely mention here that viewing two adjacent $n$-balls as
part of the cellular decomposition of an $n$-manifold, their gluing in
the sense of axiom (T5) becomes the $(n,n)$-move (or ``fusion move'')
in the terminology of \cite{Oe:tqft}.

The present framework has some similarities to so called
open-closed TQFT in two dimensions, see
\cite{LaPf:ocfrob} and references therein. It should be useful to
perform a detailed comparison between the two. For example,
open-closed TQFT also admits free boundaries that do not carry state
spaces. Such boundaries could be introduced in the present framework
by attaching fixed states to ordinary boundaries. These states would
then be seen as labels on the free boundaries.
An equivalent way to
look at this would be as Wilson loop empty regions (defined as in
Section~\ref{sec:wilson}), but with only one side.

The treatment of two-dimensional quantum Yang-Mills theory
in this work is meant merely as a first example of a quantum field
theory with corners. On the one hand, more complex theories need to be
considered. Remaining in the two-dimensional context, conformal
field theory comes to mind. On the other hand, higher dimensional
examples are of interest. The question arises in particular, whether
the axioms proposed in Section~\ref{sec:axiomlist} are ``good enough''
in higher dimensions or need to be further modified. This concerns in
particular axiom (T2), as now a whole hierarchy of dimensions comes
into play when decomposing boundaries.

A theory that should be relatively straightforward to work out is
three-dimensional quantum gravity with corners. In that case, we
merely need a topological context. The hypersurfaces will be Riemann
surfaces with holes, while the regions (general compact orientable
three-manifolds with boundary) do not admit a simple
classification. However, the latter fact should not pose any serious
problem as there is again one elementary region, the three-ball, out
of which all others can be obtained by gluing.

\subsection*{Acknowledgments}

I would like to thank J.~A.~Zapata and E.~Bianchi for stimulating
discussions. I would also like to thank I.~Runkel for carefully
reading the manuscript and alerting me to several minor mistakes.

\bibliography{stdrefs}
\bibliographystyle{amsordx}

\end{document}